\documentclass[aps,prl,twocolumn,superscriptaddress]{revtex4-2}

\usepackage{mathrsfs}
\usepackage{graphicx}
\usepackage{bm}
\usepackage{threeparttable}
\usepackage{amsmath,amssymb}
\usepackage{color}
\usepackage{epstopdf}

\begin{document}

\title{A universal scaling law for active diffusion in complex media}

\author{Qun Zhang}
\thanks{These authors contributed equally to this work.}
\affiliation{School of Physics, Beijing Institute of Technology, Beijing 100081, China}
\author{Yuxin Tian}
\thanks{These authors contributed equally to this work.}
\affiliation{School of Physics, Beijing Institute of Technology, Beijing 100081, China}
\author{Xue Zhang}
\affiliation{School of Physics, Beijing Institute of Technology, Beijing 100081, China}
\author{Xiaoting Yu}
\affiliation{School of Physics, Beijing Institute of Technology, Beijing 100081, China}
\author{Hongwei Zhu}
\affiliation{School of Physical Science and Technology, Key Laboratory of Magnetism and Magnetic Materials for Higher Education in Inner Mongolia Autonomous Region, Baotou Teachers' College, Baotou 014030, China}
\author{Ning Zheng}
\email{ningzheng@bit.edu.cn}
\affiliation{School of Physics, Beijing Institute of Technology, Beijing 100081, China}
\author{Luhui Ning}
\email{lhningphy@cup.edu.cn}
\affiliation{Beijing Key Laboratory of Optical Detection Technology for Oil and Gas, China University of Petroleum-Beijing, Beijing 102249, China}
\author{Ran Ni}
\email{r.ni@ntu.edu.sg}
\affiliation{School of Chemistry, Chemical Engineering and Biotechnology, Nanyang Technological University, 62 Nanyang Drive, Singapore 637459, Singapore}
\author{Mingcheng Yang}
\email{mcyang@iphy.ac.cn}
\affiliation{Beijing National Laboratory for Condensed Matter Physics and Laboratory of Soft Matter Physics, Institute of Physics, Chinese Academy of Sciences, Beijing 100190, China}
\affiliation{School of Physical Sciences, University of Chinese Academy of Sciences, Beijing 100049, China}
\author{Peng Liu}
\email{liupeng@bit.edu.cn}
\affiliation{School of Physics, Beijing Institute of Technology, Beijing 100081, China}

\begin{abstract}
Using granular experiments and computer simulations, we investigate the long-time diffusion of active tracers in a broad class of complex media composed of frozen obstacles of diverse structures. By introducing a dimensionless persistence length $Q = v_d \tau_r / d_t$, we propose a modified scaling relation that independently collapses experimental and simulation results across active and passive particles, diverse media, and distinct propulsion mechanisms. Our results reveal a universal active diffusion-structure relation that holds across both equilibrium and nonequilibrium regimes, providing a simple predictive framework for active diffusion in complex environments.
\end{abstract}

\pacs {}
\maketitle

\setlength{\parskip}{0pt}

\emph{Introduction.}---Active matter systems consist of self-propelled particles that exhibit either translational~\cite{Zhang2010pnas,Theurkauff2012prl,Buttinoni2013prl} or rotational motility~\cite{Noji1997Nature,Catchmark2005small,Tsai2005prl,Yang2014sm,Maggi2015nc}. Biological entities include bacteria~\cite{Berg1972nature,Wu2021nature,Zhang2024prx}, protozoa~\cite{Machemer1972jeb,Blake1974brcps}, and spermatozoa~\cite{Woolley2003reproduction,Riedel2005science}, whose propulsion is powered by molecular motor-driven flagella or cilia~\cite{Lauga2012pt,Alizadehrad2015pcb,Elgeti2015rpp}. Synthetic microswimmers, by contrast, are typically actuated by chemical reactions~\cite{Lyu2023acsnano,Wu2023nc} or external fields such as light~\cite{Zong2015acsnano,Tkachenko2023nc}, magnetic~\cite{Grzybowski2002science,Ceron2023pnas}, or acoustic stimuli~\cite{Janiak2023nc,Surappa2025nc}. Their directed motion shows promise in a wide range of applications, from targeted delivery of drugs, biomarkers, and contrast agents~\cite{Nelson2010arbe,Wang2012acsnano,Patra2013nanoscale,Abdelmohsen2014jmcb}, to the design of smart materials and microdevices~\cite{Ramaswamy2010arcmp,Vicsek2012pr,Cates2012rpp,Marchetti2013rmp,Bechinger2016rmp}.

While many of these applications inherently involve spatially heterogeneous and structural environments, such as biological tissues, porous media, and microfluidic networks, present understanding on active transport has been primarily focused on homogeneous settings. In idealized homogeneous structureless situations, the dynamics of an isolated active particle is characterized by a persistent ballistic motion at short times and an enhanced diffusion at long times due to random reorientation~\cite{Howse2007prl,Golestanian2009prl}. This crossover can be quantitatively captured by the mean square displacement (MSD), with the long-time diffusion coefficient following the well-established relation $D_a=D_t+v_d^2\tau_r/2$, where $v_d$ is the propulsion speed and $\tau_r$ the rotational relaxation time~\cite{Franke1990ebj,Howse2007prl,Martens2012epje}.

In contrast, it is known that complex background structure can drastically alter dynamical behavior of active particles, giving rise to phenomena such as subdiffusion and trapping~\cite{Chepizhko201310prl,Takagi2014sm,Spagnolie2015sm,Sipos2015prl,Chepizhko2015epjst}, dynamic freezing~\cite{Volpe2014ajp,Volpe2014oe,Paoluzzi2014jpcm,Pesce2015josab}, and chirality-dependent sorting~\cite{Nourhani2015prl,Schirmacher2015prl,Ai2015sm,Chen2015jcp}. These observations raise a fundamental question: how does the structure of complex environments govern active diffusion? And further whether a general active diffusion-structure relation exists? Exploring a possible universal relation between passive tracer diffusion and background structure has attracted considerable interest and made great progress in recent decades~\cite{Dzugutov1996nature,Rosenfeld1999jpcm,Samanta2004prl,Ning2019prl,Ning2021pre}, however the exploration of its active counterpart is still lacking.

In this Letter, combining granular experiments and particle-based simulations, we investigate the diffusion of small active particles in various spatially heterogeneous environments consisting of obstacles of disordered liquid-like structures, or ordered square and hexagnal crystal lattices. We fabricate 3D-printed polylactide particles featuring tilted legs that allow the activity tuning via the inclination angle~\cite{Scholz2016njp,Scholz2017prl,Scholz2018nc,Altshuler2013po}. In simulations, self-propulsion is modeled using active Brownian motion~\cite{Howse2007prl,Palacci2013science,Liu2020prl}, run-and-tumble dynamics~\cite{Berg1979nature,Tailleur2008prl}, and active Ornstein–Uhlenbeck processes~\cite{Uhlenbeck1930pr,Wu2000prl,Fodor2016fprl}, both overdamped and underdamped. We find that the normalized diffusion coefficient $D^*$, incorporating activity, can be collapsed onto a single master curve as a function of the two-body structural entropy $S_2$~\cite{Ma2013prl,Thorneywork2015prl,Ning2019prl}, which is independent of the self-propulsion mechanism or the medium structure. This finding reveals a general active diffusion-structure relation and provides a predictive framework for long-time active diffusion in spatially heterogeneous environments.

\begin{figure}[t]
\includegraphics[width=0.47\textwidth]{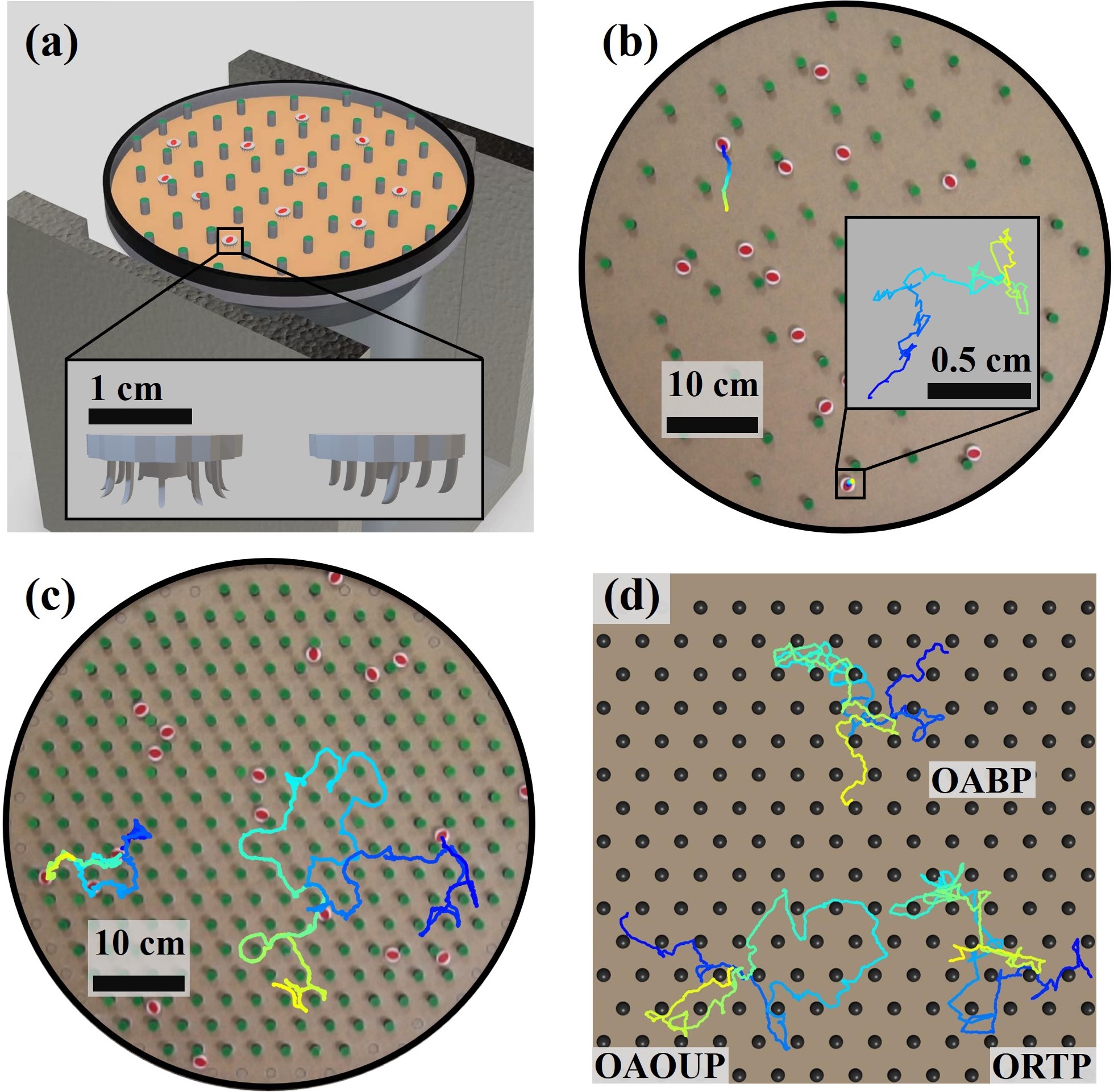}
\caption{(a) Schematic of the quasi-2D electromagnetic shaker setup. The inset highlights 3D-printed tracer particles with symmetric (passive) and asymmetric (active) legs; propulsion speed $v_d$ is tuned via the leg inclination angle. (b) Short-time trajectories ($0$-$0.7$~s) in a disordered background, where the upper path corresponds to an active tracer ($v_d=4.08$~cm/s) and the lower path to a passive particle undergoing thermal diffusion. (c) Long-time trajectories ($0$-$70$~s) in a hexagonal lattice: left, passive particle; right, active tracer (same as in (b)). (d) Simulated long-time trajectories of OABP, OAUP, and ORTP tracers in a hexagonal lattice over $5\times10^5$ steps. In (b)-(d), trajectory color indicates time, from blue (start) to yellow (end).}\label{Fig1}
\end{figure}

\emph{Experiment.}---Disk-like tracers were placed in a $54$-cm-diameter circular vessel mounted on an electromagnetic shaker [Fig.~\ref{Fig1}(a)]. Fixed polymethyl methacrylate (PMMA) cylinders with diameter $d_l=1$ cm and height $h=2$ cm served as immobile obstacles, arranged in three types of backgrounds: disordered (number density $0.005$-$0.025$ cm$^{-2}$), square and hexagonal lattices, both with the lattice constant $L=3$-$8$ cm. Vertical vibrations were applied with a magnitude of $Z=\zeta\sin(2\pi ft)$ and frequency $f=80$ Hz. The dimensionless acceleration $\tilde{\Gamma}_{vib}=\zeta(2\pi f)^2/g=3.0$, with $g$ the gravitational acceleration, maintaining stable quasi-2D dynamics.

Tracers with diameter $d_t=1.5$ cm were 3D-printed from polylactide, equipped with either symmetric (passive) or asymmetric (active) legs [inset, Fig.~\ref{Fig1}(a)]~\cite{Scholz2016njp,Scholz2017prl,Scholz2018nc,Altshuler2013po}. For active tracers, the asymmetric leg geometry rectifies vertical vibrations into persistent horizontal motion. Propulsion speeds $v_d = 2.04$, $2.72$, $3.40$, and $4.08$ cm/s were tuned via the leg inclination angle~\cite{Koumakis2016njp}. Particle trajectories were extracted using particle tracking techniques~\cite{Crocker1996jcis,Ning2023prl}, excluding regions within $1/8$ of the vessel diameter to avoid boundary effects. As shown in Figs.~\ref{Fig1}(b) and (c), passive tracers undergo normal diffusion, while active tracers exhibit a superdiffusive-to-diffusive crossover, consistent with the persistent active Brownian motion [see the movies in the Supplementary Material].

\emph{Simulation.}---Numerical simulations were performed in a 2D periodic box containing $1600$ immobile matrix particles arranged in disordered, square, and hexagonal configurations, matching the experiments. Under the standard reduced simulation units, the radius of matrix particles and thermal energy $k_{\tt B}T$ are taken as the units of length and energy, respectively. The tracer diameter $d_t$ is $25\%$ of the matrix particle diameter $d_l$. Tracer-matrix interactions were modeled by a truncated repulsive Lennard-Jones potential, $U(r)=4\epsilon\left[\left(\frac{\sigma}{r}\right)^{2n}-\left(\frac{\sigma}{r}\right)^{n}\right]+\epsilon$, $r<2^{1/n}\sigma$, with the interaction diameter $\sigma=(d_t + d_l)/2$, interaction strength $\epsilon=k_{\tt B}T=1.0$, and stiffness $n = 12$. Matrix packing fractions were varied to access a wide range of structural entropies. Active tracers propelled with an effective diffusivity $D_a=D_t+v_d^2\tau_r/2$ and reorient via three representative models: active Brownian particles (ABPs), run-and-tumble particles (RTPs), and active Ornstein–Uhlenbeck particles (AOUPs), which capture the stochastic dynamics of synthetic Janus colloids~\cite{Howse2007prl,Palacci2013science,Liu2020prl}, \emph{E.~coli}-like swimmers~\cite{Berg1979nature,Tailleur2008prl}, and particles driven by active baths~\cite{Uhlenbeck1930pr,Wu2000prl,Fodor2016fprl}, respectively. Each model was implemented in both overdamped and underdamped forms, labeled as OABP, ORTP, OAOUP and UABP, URTP, UAOUP, respectively. Translational and rotational thermal noises were included as independent Gaussian-distributed forces and torques following the fluctuation-dissipation relation. The persistence length $v_d \tau_r/d_t$ reached up to $6$, significantly exceeding the experimental maximum of $1.9$ limited by vibrational constraints. Representative long-time trajectories are shown in Fig.~\ref{Fig1}(d). Simulation details are provided in the Supplementary Material.

\begin{figure}[t]
\includegraphics[width=0.47\textwidth]{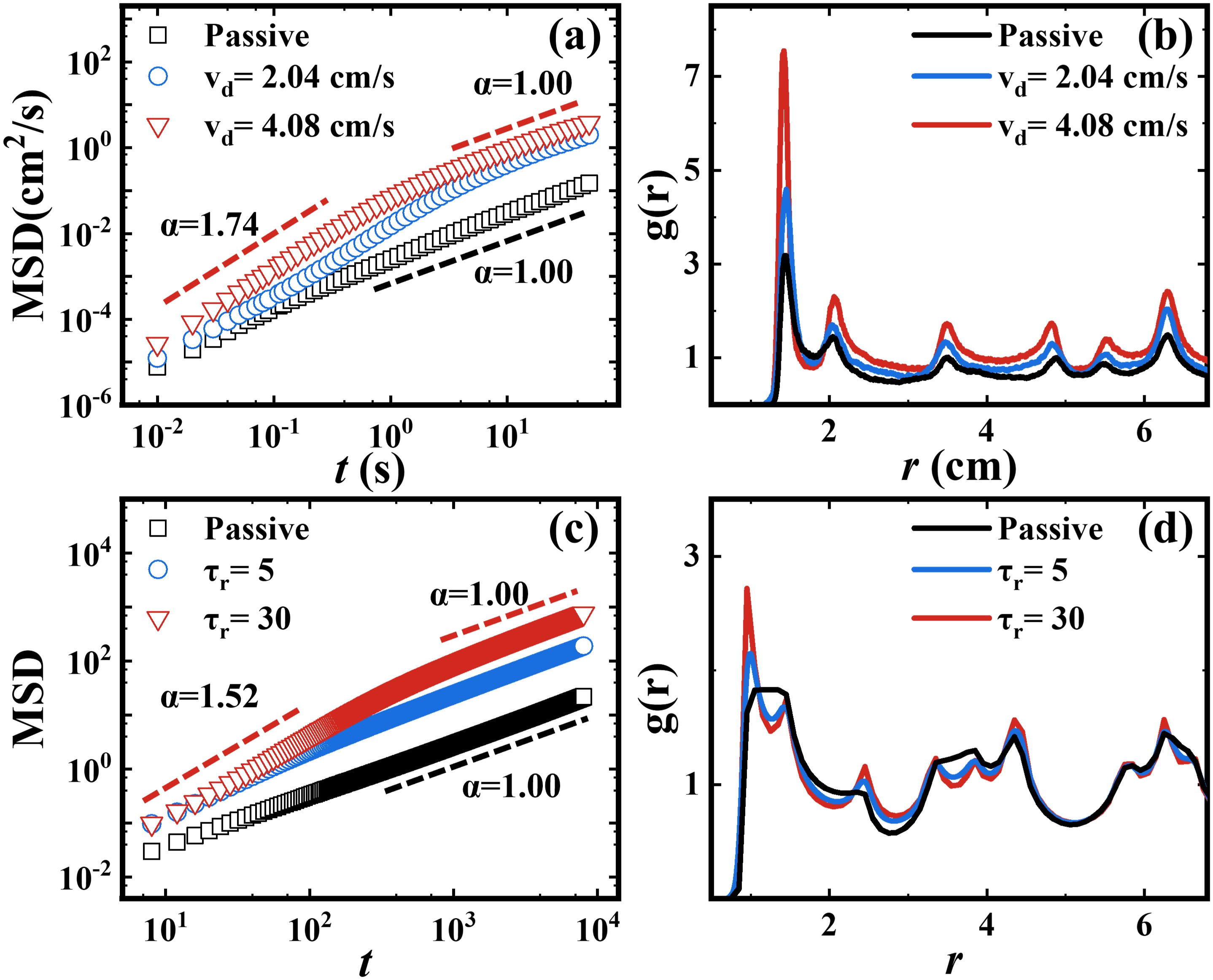}
\caption{(a) MSDs as a function of time $t$, and (b) the radial distribution function $g(r)$ between tracers and background particles for passive and active tracers with propulsion speeds $v_d=2.04$~cm/s and $4.08$~cm/s, measured experimentally in a hexagonal lattice with lattice constant $L=3$~cm. (c) and (d) Corresponding MSD and $g(r)$ from simulations in a square lattice at packing fraction $\rho=0.55$, for passive tracers and active tracers with rotational relaxation times $\tau_r=5$ and $30$.}\label{Fig2}
\end{figure}

\emph{Results and discussion.}---In both experiments and simulations, the long-time dynamics of active tracers are characterized by the diffusion coefficient $D_a$, extracted from MSD defined as $\langle\Delta\mathbf{r}^2(t)\rangle=\langle[\mathbf{r}(t_0+t)-\mathbf{r}(t_0)]^2\rangle$, where $\langle\cdot\rangle$ denotes the ensemble average. Figure~\ref{Fig2}(a) shows log-log plots of MSDs for passive and active tracers in a hexagonal lattice ($L = 3$ cm) in experiments. Active tracers exhibit a crossover from the short-time superdiffusion ($\langle\Delta\mathbf{r}^2(t)\rangle\propto t^\alpha$, $\alpha > 1$) to the long-time normal diffusion ($\alpha = 1$). The long-time diffusion coefficient $D_a$ is extracted by fitting the MSD with $\langle \Delta\mathbf{r}^2(t) \rangle=4D_at$. Simulated MSDs at slightly higher matrix packing fractions are shown in Fig.~\ref{Fig2}(c), with $D_a$ obtained using the same fitting procedure.

The structure of the system is characterized by the two-body structural entropy $S_2$, which approximates the excess entropy, \emph{i.e.}, the entropy difference between a correlated system and an ideal gas under identical conditions. While the full excess entropy includes higher-order correlations, the two-body contribution $S_2$ accounts for more than $85\%$ of the total excess entropy across a broad density range, and has been widely used as an approximation in complex systems~\cite{Baranyai1989pra}. For a tracer in a matrix, $S_2$ is computed from the pair correlation function $g(r)$ between the tracer and background particles~\cite{Ma2013prl,Thorneywork2015prl,Ning2019prl}:
\begin{equation}
S_2=-\pi\rho\int_0^{\infty}\{{g(r)\ln[g(r)]-[g(r)-1]}\}rdr,\label{eq1}
\end{equation}
where $\rho$ is the number density of background matrix particles. Given the low tracer concentration, tracer-tracer interactions are normally negligible, and only tracer-matrix correlations are considered. The function $g(r)$, defined as the probability of finding a matrix particle at distance $r$ from a tracer, is measured by
\begin{equation}
g(r)=\frac{1}{2\pi r \rho} \left\langle \frac{1}{N_t} \sum_{i=1}^{N_t} \sum_{j=1}^{N_l} \delta(r - r_{ij}) \right\rangle, \label{eq2}
\end{equation}
where $N_t$ and $N_l$ are the numbers of tracers and matrix particles, respectively, and $r_{ij}$ is the distance between tracer $i$ and matrix particle $j$. Experimentally measured $g(r)$ in a hexagonal lattice is shown in Fig.~\ref{Fig2}(b). As the propulsion speed $v_d$ increases from $0$ (passive) to $4.08$ cm/s (active), the first peak of $g(r)$ increases, indicating stronger tracer-matrix correlations. Periodic peaks in $g(r)$ arise from the crystalline order of the matrix and are similarly observed in simulations at varying $\tau_r$ [Fig.~\ref{Fig2}(d)].

\begin{figure}[b]
\includegraphics[width=0.38\textwidth]{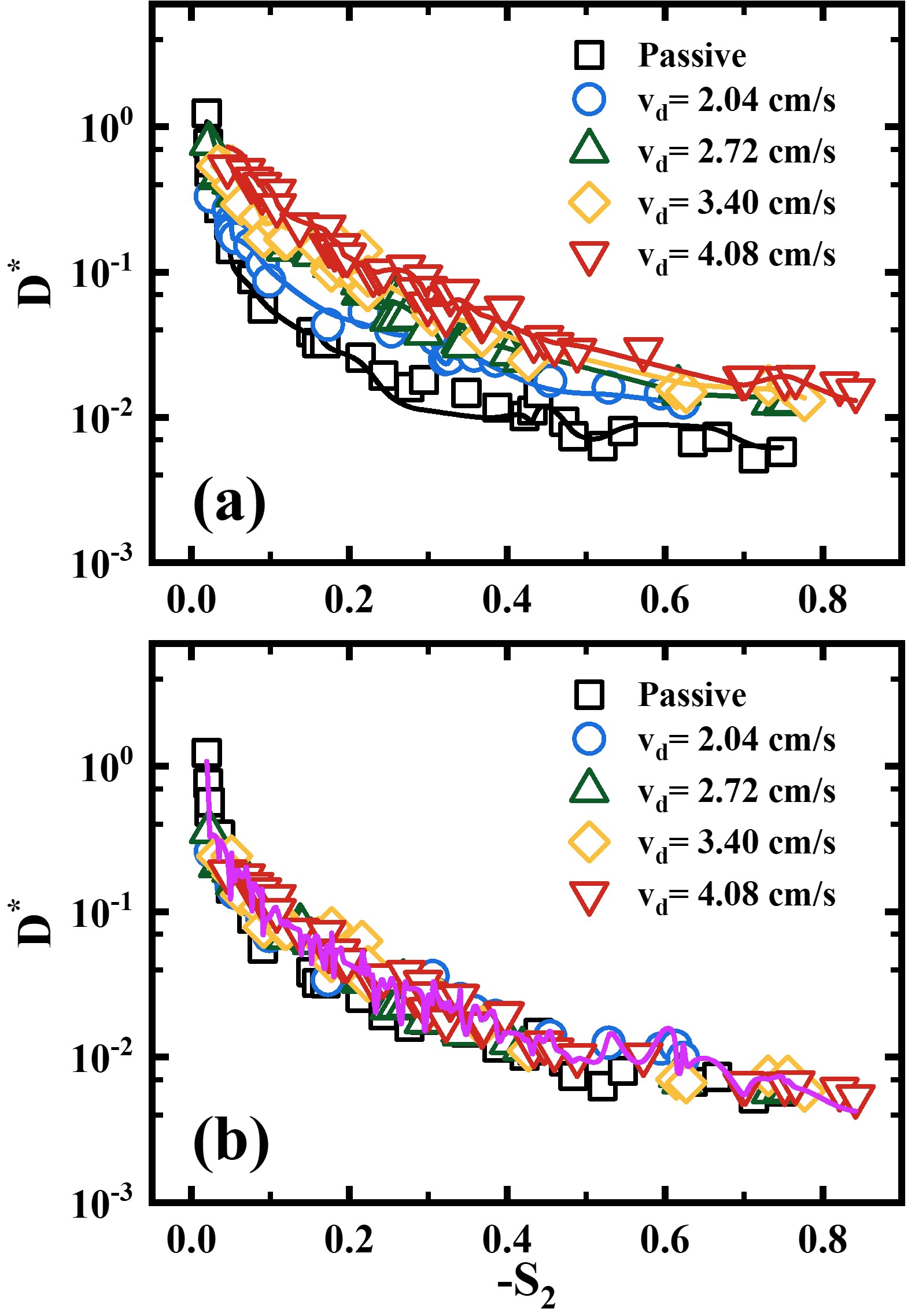}
\caption{(a) Experimentally measured dimensionless diffusion coefficient $D^*=D\Gamma^{-1}d_t^{-2}$ as a function of two-body structural entropy $S_2$ for passive and active tracers with varying self-propulsion speeds $v_d$ in disordered and ordered backgrounds. Solid lines show fits to the original scaling relation Eq.~\ref{eq4}. (b) Same data replotted using the modified normalized form $D^*=D\left[\Gamma(1+Q)\right]^{-1}d_t^{-2}$ that incorporates activity. All data collapse onto a single master curve described by improved scaling relation Eq.~\ref{eq5}, capturing a unified trend across both passive and active tracers.}\label{Fig3}
\end{figure}

\begin{figure}[t]
\includegraphics[width=0.38\textwidth]{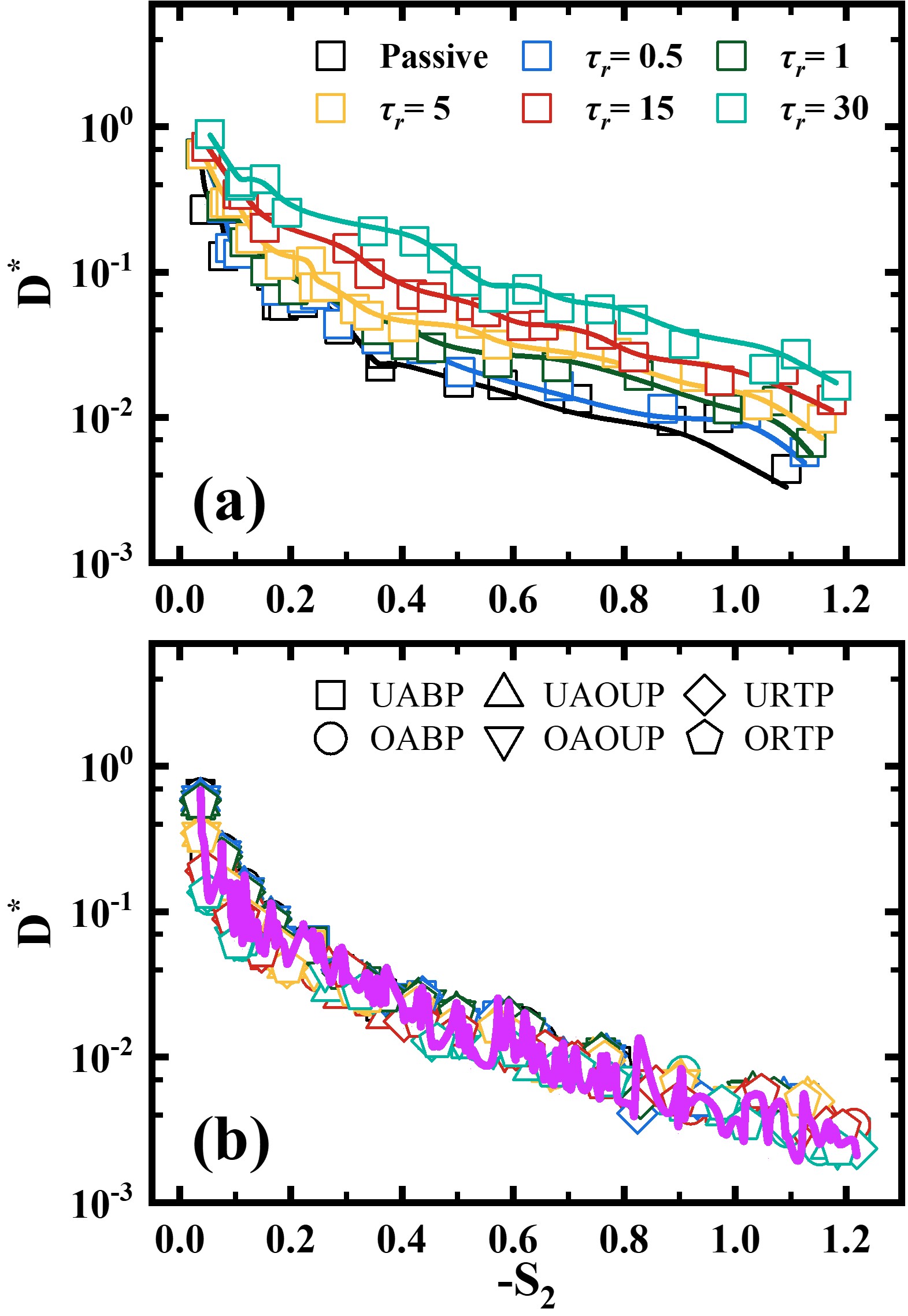}
\caption{(a) Simulated dimensionless diffusion coefficient $D^*=D\Gamma^{-1}d_t^{-2}$ as a function of $S_2$ for passive and active tracers in the UABP model in disordered and ordered backgrounds. Solid lines show fits to the original scaling relation Eq.~\ref{eq4}. Colored markers represent active particles, with rotational relaxation time $\tau_r$ increasing from $0.5$ to $30$. (b) Same data replotted using the modified normalization $D^*=D\left[\Gamma(1+Q)\right]^{-1}d_t^{-2}$, along with results from five additional active motion models over the same $\tau_r$ range. All data collapse onto a single master curve described by the improved scaling relation Eq.~\ref{eq5}, capturing a unified trend across passive and active tracers and across diverse microscopic reorientation dynamics.}\label{Fig4}
\end{figure}

In the following, we normalize the long-time diffusion coefficients---both passive ($D_t$) and active ($D_a$)---using the dimensionless form $D^*=D\Gamma^{-1}d_t^{-2}$~\cite{Ning2019prl,Ning2021pre}. The Enskog collision frequency $\Gamma$ between the tracer and immobile matrix particles in 2D is given by~\cite{Wang2015pre,Ning2019prl}
\begin{equation}
\Gamma=2\sigma\rho g(\sigma)\sqrt{\pi k_{\tt B}T_k/m}, \label{eq3}
\end{equation}
where $\sigma$ is the position of the first peak in $g(r)$, $\rho$ is the matrix number density, and $m$ is the reduced mass, defined as $m=2m_t m_l / (m_t+m_l)$, with $m_t$ and $m_l$ the masses of the tracer and background particles, respectively. Since matrix particles are fixed, they effectively form an infinite-mass rigid background, and we take the limit $m_l\to\infty$, yielding $m\approx 2m_t$. The kinetic temperature $k_{\tt B}T_k$ is extracted from the average kinetic energy of passive tracers in the vibrated system without obstacles~\cite{McNamara1998pre,Tatsumi2009jfm}. For active tracers, $k_{\tt B}T_k$ is scaled by $D_a/D_t$ to account for their enhanced long-time diffusivity under constant translational damping.

Figure~\ref{Fig3}(a) plots the normalized diffusion coefficients $D^*$ as a function of the two-body structural entropy $S_2$ for tracers with varying activity in disordered and ordered backgrounds. The experimental data do not collapse onto a single curve; instead, each activity level follows its own trend, albeit broadly consistent with the previously proposed structure-dynamics scaling relation~\cite{Ning2019prl,Ning2021pre}:
\begin{equation}
D^*=\frac{k_{\tt B}T_k}{\Gamma d_t^2\xi_B}\frac{1}{A(1-S_2)+\xi_S/\xi_B}, \label{eq4}
\end{equation}
where $\xi_B=2\sigma\rho g(\sigma)\sqrt{\pi k_{\tt B}T_km}$ is the effective friction due to tracer-matrix collisions~\cite{Longuet1956jcp,Garcia2006pre,Srivastava2009jced}, and $A$ is a fitting parameter. The friction coefficient from the substrate, $\xi_S$, is determined by using the Einstein relation for passive diffusion in the absence of obstacles, $\xi_S=k_{\tt B}T_k/D_t$, an assumption widely adopted in granular systems~\cite{Makse2002nature,Abate2008prl,Wang2022prl}. This confirms that long-time diffusivity is largely controlled by the two-body structural entropy $S_2$ of the environment. However, $D^*$ systematically increases with tracer activity even at fixed $S_2$, and fitting Eq.~\ref{eq4} requires activity-dependent values for the fitting parameter $A$. This suggests that while Eq.~\ref{eq4} holds for subsets of fixed activity, it fails to quantify the universal nonequilibrium dynamics.

In particular, the kinetic temperature $k_{\tt B}T_k$ in the collision frequency $\Gamma$ [Eq.~\ref{eq3}] accounts only for the enhanced diffusivity but neglects the directional persistence unique to active motion. Unlike passive tracers, active particles tend to maintain contact with obstacles upon collision rather than being scattered instantly, which effectively amplifies the collision frequency. To capture this, we introduce a dimensionless persistence length $Q = v_d \tau_r / d_t$, which increases with propulsion speed and reorientation time. This quantity characterizes the directional memory that enhances tracer-obstacle coupling beyond purely thermal models. We then define a modified dimensionless diffusion coefficient as $D^*=D\left[\Gamma(1+Q)\right]^{-1}d_t^{-2}$, which recovers the passive case in the limit of $Q = 0$. Substituting this into Eq.~\ref{eq4} leads to the generalized scaling form:
\begin{equation}
D^*=\frac{k_{\tt B}T_k}{\Gamma(1+Q)d_t^2\xi_B}\frac{1}{A(1-S_2)+\xi_S/\xi_B}. \label{eq5}
\end{equation}
The self-propulsion speed $v_d$ is extracted from particle trajectories, and the rotational relaxation time $\tau_r$ is determined from $D_a=D_t+v_d^2\tau_r/2$ in the absence of obstacles. As shown in Fig.~\ref{Fig3}(b), all experimental data---spanning varying propulsion speeds and matrix structures---collapse onto a single master curve, in quantitative agreement with the modified scaling relation Eq.~\ref{eq5} with a single fitting parameter $A=1$. This indicates that $D^*$ is entirely determined by $S_2$ and $Q$, with no discontinuities observed across the transition from passive to active diffusion. This establishes a unified scaling framework for long-time diffusion in heterogeneous media, bridging equilibrium and nonequilibrium systems via two experimentally accessible parameters.

Next, we further test the universality of the scaling relation through simulations of active colloidal tracers with diverse reorientation dynamics. Three representative models ---ABP, AOUP, and RTP--- are considered, each implemented under both underdamped and overdamped conditions to assess the effect of inertia. Figure~\ref{Fig4}(a) plots the dimensionless diffusion coefficient $D^*$ as a function of $S_2$ for passive and active tracers in the UABP model. Like in Fig.~\ref{Fig3}(a), data points with different activities remain dispersed and do not collapse onto a single curve. Each trend is described by the original scaling relation Eq.~\ref{eq4}, with an activity-dependent fitting parameter $A$. Passive tracers, shown in black, serve as a reference for comparison. Results for the other five active motion models are presented in the Supplementary Material. Remarkably, when the same data are replotted using the modified normalization in Fig.~\ref{Fig4}(b), all simulation results (across reorientation mechanisms, damping regimes, and matrix structures) collapse onto a single universal curve, in excellent agreement with the modified scaling relation, Eq.~\ref{eq5}, with a single parameter $A=8$ independent of active models. The larger fitting parameter compared to the experimental value can be attributed to the fact that in macroscopic granular experiments, the Einstein relationship is assumed, while particle-obstacle collisions are inelastic, and the noise is athermal. This demonstrates that the long-time diffusion of active tracers in complex media is governed entirely by the two-body structural entropy $S_2$ and the dimensionless persistence length $Q = v_d \tau_r / d_t$, regardless of the underlying stochastic process.

\emph{Conclusion.}---Our work establishes a universal scaling law for active diffusion in complex media, governed by two parameters: the structural entropy $S_2$ and the dimensionless persistence length $Q = v_d \tau_r / d_t$. Here, $S_2$ quantifies the environmental constraints arising from structural correlations, while $Q$ captures the persistence of active propulsion. This finding paves the way towards understanding, controlling and exploiting the diffusive transportation of active particles in complex environment.

\section*{Acknowledgment}
We acknowledge the supports of the National Natural Science Foundation of China (Grants No. 12374205, No. 12304245, No. T2325027, No. 12274448, and No. 12475031), and the Natural Science Foundation of Shandong Province (Grant No. ZR2024YQ017). This work was also supported by Beijing National Laboratory for Condensed Matter Physics (Grants No. 2023BNLCMPKF014, and No. 2024BNLCMPKF009), the Academic Research Fund from the Singapore Ministry of Education (RG59/21), and the National Research Foundation, Singapore, under its 29th Competitive Research Program (CRP) Call (Award No. NRF-CRP29-2022-0002).


\end{document}